\documentclass[a4paper,11pt]{article}
\usepackage{pos}
\usepackage{subcaption}

\title{Thermal interquark potentials for bottomonium using NRQCD from the HAL QCD method}

\ShortTitle{Thermal interquark potentials for bottomonium}
\author*[a]{Thomas Spriggs}
\author[a]{Chris Allton}
\author[a]{Timothy Burns}
\author[b]{Seyong Kim}

\affiliation[a]{Department of Physics, Swansea University, Swansea SA2 8PP, United Kingdom}

\affiliation[b]{Department of Physics, Sejong University, Seoul 143-747, Korea}

\emailAdd{t.spriggs.996870@swansea.ac.uk}

\abstract{ We report our preliminary progress in the calculation of
  the interquark potential of bottomonium at non-zero temperature
  using the HAL QCD method. We use NRQCD correlation functions of
  non-local mesonic $S$-wave states to obtain the central potential as a
  function of temperature.  These results have been obtained using our
  anisotropic 2+1 flavour "Generation 2" FASTSUM ensembles.}

\FullConference{%
 The 38th International Symposium on Lattice Field Theory, LATTICE2021
  26th-30th July, 2021
  Zoom/Gather@Massachusetts Institute of Technology
}

\begin{document}
\maketitle

\section{Introduction}

The interquark potential of quarkonia is one of the first quantities
studied in the quest for a deeper understanding of the nature of the
strong interaction.  Pioneering studies include \cite{Eichten:1978tg}
where the Cornell potential was used to calculate the spectrum of
charmonium states using a Quantum Mechanical
formalism.  In thermal QCD, the temperature dependence of the interquark potential results in quarkonium states melting at
different temperatures \cite{Matsui:1986dk}.  These considerations
strongly motivate a study of the thermal behaviour of the quarkonia
interquark potential.

Heavy quarks interacting via QCD can be approximated using the
non-relativistic approach, NRQCD, which allows a significant
simplification. NRQCD calculations of bottomonia are typically
accurate at the few percent level.  In this work we use NRQCD to
determine the interquark potential in bottomonia using the HAL QCD
approach. Correlation functions of bottomonia operators are studied
where the quark and antiquark are non-local, and this allows a proxy
for the wavefunction to be calculated. Using this wavefunction in the
Schr\"odinger equation leads to the interquark potential.
We find indications of the weakening of the potential
as the temperature increases, as expected.
This work extends previous studies of the interquark potential
by the FASTSUM Collaboration in the charmonium system \cite{Evans:2013yva,Allton:2015ora}.

\section{NRQCD correlation functions and lattice setup}

NRQCD is an effective theory with a power counting in the heavy quark
velocity, $v$. In this theory, the heavy quark and antiquark fields
decouple and so virtual heavy quark-antiquark loops cannot form. The NRQCD
quark propagator is calculated via an initial value problem, rather
than via a matrix inversion as is the case for relativistic quarks.
NRQCD is particularly amenable for lattice simulations because mesonic
correlation functions do not have ``backward movers'' which complicate
the study of QCD mesons.

Our NRQCD formulation incorporates both ${\cal O}(v^4)$ and the leading
spin-dependent corrections. The $b$-quark mass is tuned by setting the
``kinetic'' mass (i.e. from the dispersion relation) of the
spin-averaged $1S$ states to its experimental value.  Full details of
our NRQCD setup appear in \cite{Aarts:2014cda}.

All our results were obtained using our FASTSUM $N_f=2+1$ flavour
``Generation 2'' ensembles which have the parameters listed in Table
\ref{tab:fastsum_details}.

\begin{table}[h!]
    \centering
        \begin{tabular}{c||c|c|c|c|c|c|c}
        $N_\tau$ & 16 & 20 & 24 & 28 & 32 & 36 & 40 \\
        \hline
        T [MeV] & 352 & 281 & 235 & 201 & 176 & 156 & 141 \\
        \hline
        $N_{\text{configurations}}$ & 1050 & 950 & 1000 & 1000 & 1000 & 500 & 500 
    \end{tabular}
    \caption{An overview of the FASTSUM Generation 2 correlation
      functions used in this work. Lattice volumes are $(24 a_s)^3\times (N_\tau
      a_\tau)$ with $a_s = 0.1227(8)$fm and $a_\tau =
      35.1(2)$am. For these ensembles with a pion mass of $M_\pi = 384(4)$MeV,
      the pseudo-critical temperature T$_{\rm pc} = 181(1)$MeV \cite{Aarts:2019hrg}. }
    \label{tab:fastsum_details}
\end{table}

\section{HAL QCD Method} \label{sec:hal_qcd}

We follow the HAL QCD time-dependent method to extract the interquark potential \cite{Ishii:2006ec,Ishii:2012ssm}.
A key quantity for the HAL QCD method is the Nambu Bethe Salpeter (NBS)
wave function, $\psi_i(\textbf{r}) = \langle 0 | J(\textbf{r}) |i \rangle$,
i.e. the overlap of the non-local mesonic operator $J(\textbf{r})$
between the vacuum and the bottomonium state $|i\rangle$.  The mesonic
operator $J(\textbf{r})$ is defined
\[
J_\Gamma(x;\textbf{r}) =
\overline{Q}(x) \,U(x,x+\textbf{r})\, \Gamma Q(x+\textbf{r})
\]
and thus probes the bottomium state with a displacement of
$\textbf{r}$ between its two constituent quarks. $\Gamma$ is a Dirac
matrix chosen to have the desired quantum numbers appropriate for
either the $\Upsilon$ $(\Gamma = \gamma_i)$ or $\eta_b$ $(\Gamma =
\gamma_5)$ states, and $U(x,x+\textbf{r})$ is the gauge connection
between $x$ and $x+\textbf{r}$.

We calculate the zero-momentum correlation function
\[
G_\Gamma(\textbf{r},\tau) =
\sum_{\textbf{x}} \langle J_\Gamma        (\textbf{x},\tau; \textbf{r})
                          J_\Gamma^\dagger(0;               \textbf{0}) \rangle
= \sum_i \frac{\psi_i(\textbf{r})\psi_i^*(\textbf{0})}{2E_i} e^{-E_i\tau}
= \sum_i \Psi_i(\textbf{r})e^{-E_i\tau}.
\]
The sum over states $i$ is the usual spectral representation,
and for convenience we've defined
\[
\Psi_i(\textbf{r}) = \frac{\psi_i(\textbf{r})\psi_i^*(\textbf{0})}{2E_i}.
\]

Since we are treating the bottom quark nonrelativistically, we can
assume that $\Psi_i(\textbf{r})$ obeys the time independent
Schrodinger equation in Euclidean space-time,
\[
    \bigg(-\frac{\nabla^2_r}{2\mu} + V_\Gamma(r)\bigg)\Psi_i(\textbf{r}) = E_i\Psi_i(\textbf{r}),
\]
where $V_\Gamma(r)$ is the interquark potential for the channel
$\Gamma$, $\mu$ is the reduced mass, and we restrict to S-wave states.
Since the correlation function, $G(\tau)$, is a linear combination of
$\Psi_i(\textbf{r})$, we find that it satisfies the Schr\"odinger equation,
\begin{equation}
    \bigg(-\frac{\nabla^2_r}{2\mu} + V_\Gamma(r)\bigg)
    G_\Gamma(\textbf{r},\tau) = -\frac{\partial G_\Gamma(\textbf{r},\tau)}{d\tau}.
\label{eq:halqcd}
\end{equation}
We use eq\eqref{eq:halqcd} to extract the potential, $V_\Gamma$ 
from $G_\Gamma(\textbf{r},\tau)$.
We note that the NRQCD case considered here has a particularly simple
form because there are no backward movers. This contrasts with the
relativistic case where there are backward movers which need to be considered
\cite{Evans:2013yva,Allton:2015ora}.

We use finite derivatives to approximate the Laplacian and the temporal derivative.
Because we consider $S$-wave states with rotational symmetry, the
Laplacian in spherical coordinates can be approximated by
\begin{equation}
    \nabla_r^2f(r) = \frac{\partial^2 f}{\partial r} + \frac{2}{r} \frac{\partial f}{dr}
\approx \bigg(\frac{f(r+a_s)-2f(r)+f(r-a_s)}{a_s^2} + \frac{f(r+a_s)-f(r-a_s)}{ra_s}\bigg).
\label{eq:s-der}
\end{equation}
The time derivative is similarly approximated by
\begin{equation}
    \frac{\partial f}{\partial \tau} \approx \bigg(\frac{f(\tau+a_\tau)-f(\tau-a_\tau)}{2a_\tau}\bigg).
\label{eq:t-der}
\end{equation}

Using the leading order terms in the velocity expansion of the interquark potential for $S$-wave states \cite{Godfrey:1985xj},
the central potential can be defined in terms of the potential from the pseudoscalar (i.e. the $\eta_b$) and vector ($\Upsilon$) channels' potentials,
\begin{equation}
    V_c(r) = \frac{1}{4}V_{\text{PS}}(r) + \frac{3}{4}V_{\text{V}}(r).
    \label{eq:Vcentral}
\end{equation}
The spin-dependent potential is also accessible to us from these two channels, but is
not considered here. Higher order terms in the potential such as the spin-orbit term
are also not studied because they require channels with orbital angular momentum.

\section{Results}

\subsection{Derivatives}

\begin{figure}[h]
\begin{subfigure}{.5\textwidth}
  \centering
    \includegraphics[width=\linewidth]{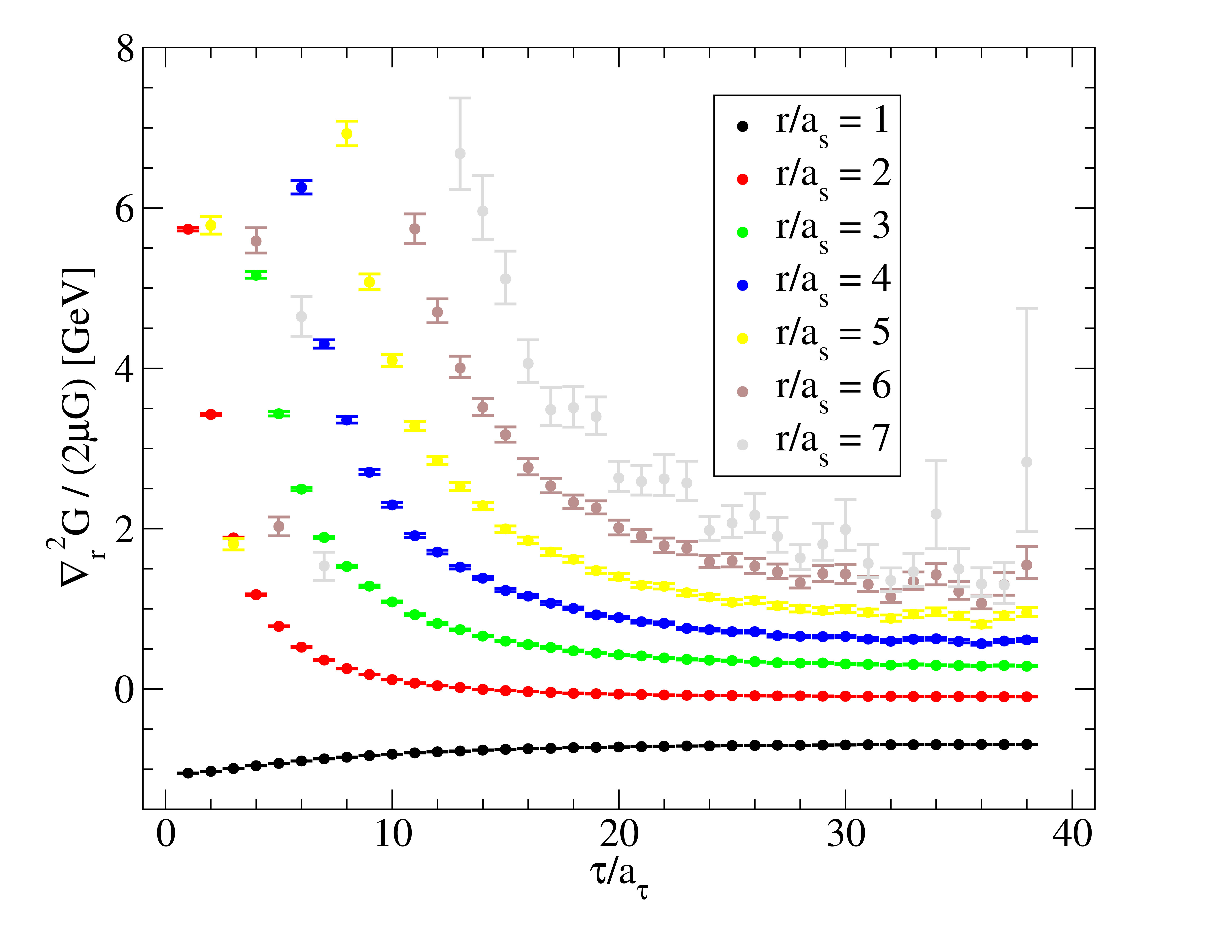}
\end{subfigure}
\begin{subfigure}{.5\textwidth}
  \centering
    \includegraphics[width=\linewidth]{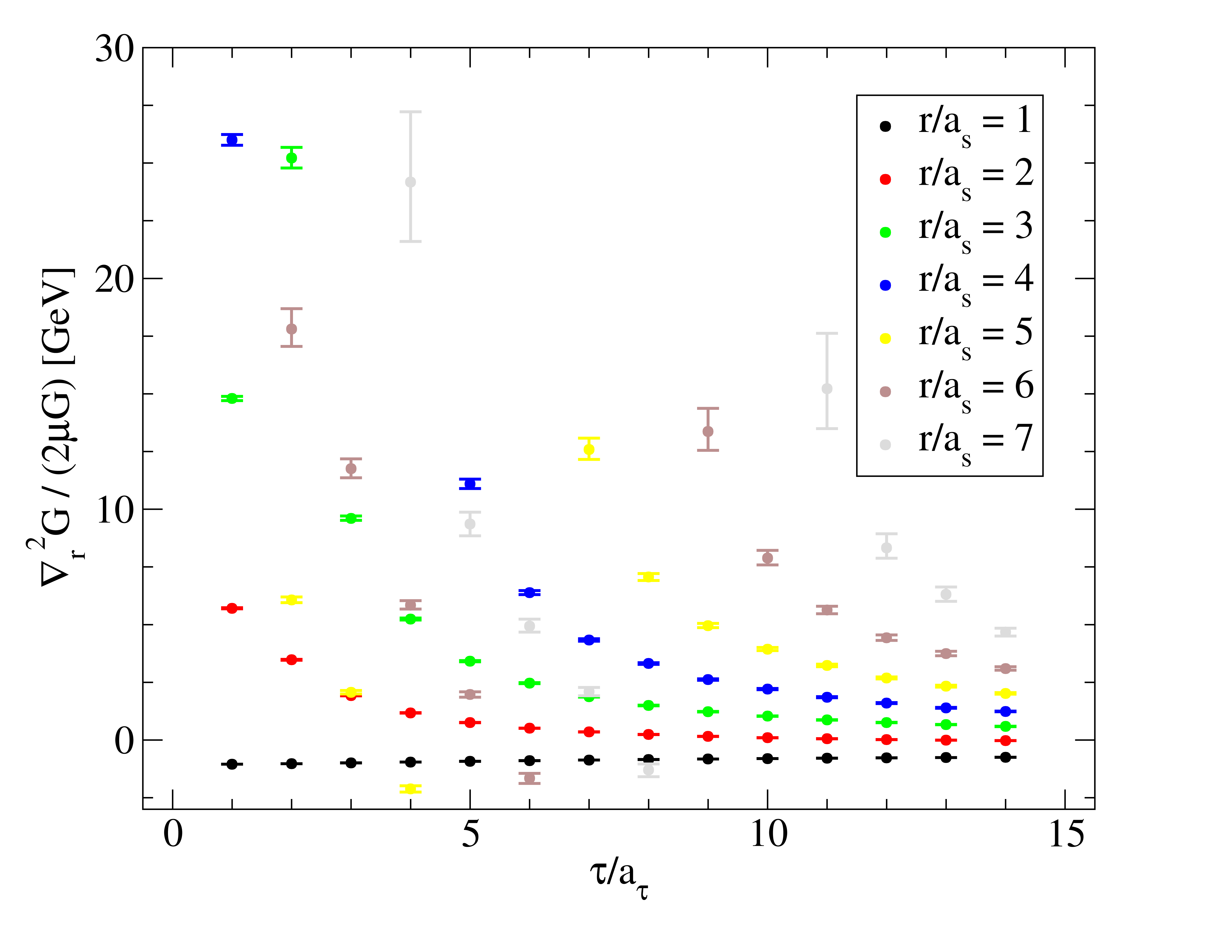}
\end{subfigure}
    \caption{The kinetic (i.e. spatial derivative) contribution to the $\Upsilon$ potential plotted against imaginary time for $T=141$MeV (left) and $T=352$MeV (right). For both temperatures the noise increases with distance, $r/a_s$. The $T=352$MeV data shows larger variation and the axis scale is adjusted to reflect this. }
    \label{fig:spatial_deriv_ntau40}
\end{figure}

We begin by separately studying the spatial and temporal derivatives in
eq\eqref{eq:halqcd}.
Figure \ref{fig:spatial_deriv_ntau40} shows the spatial derivative,
i.e. the kinetic contribution for
the $\Upsilon$ at two indicative temperatures, $T=141$MeV (left) and $T=352$MeV (right). 
For both temperatures the $\tau$ dependence becomes more noticeable at larger $r$,
as does the size of the statistical errors. The increase in noise at larger $r$
is to be expected for point-split lattice correlation functions since points close
together are correlated and so fluctuations increase with $r$.

\begin{figure}[h]
\begin{subfigure}{.5\textwidth}
  \centering
    \includegraphics[width=\linewidth]{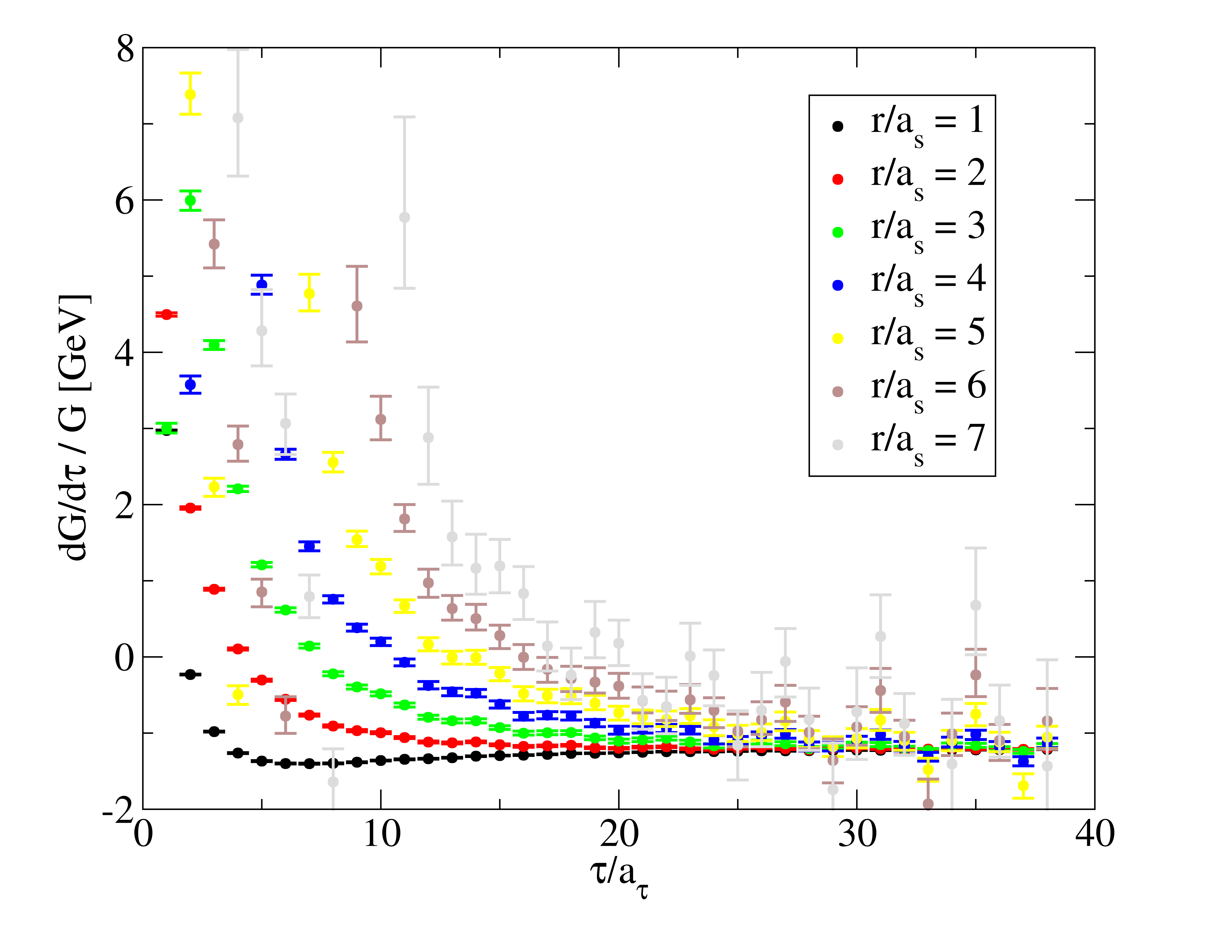}
\end{subfigure}
\begin{subfigure}{.5\textwidth}
  \centering
    \includegraphics[width=\linewidth]{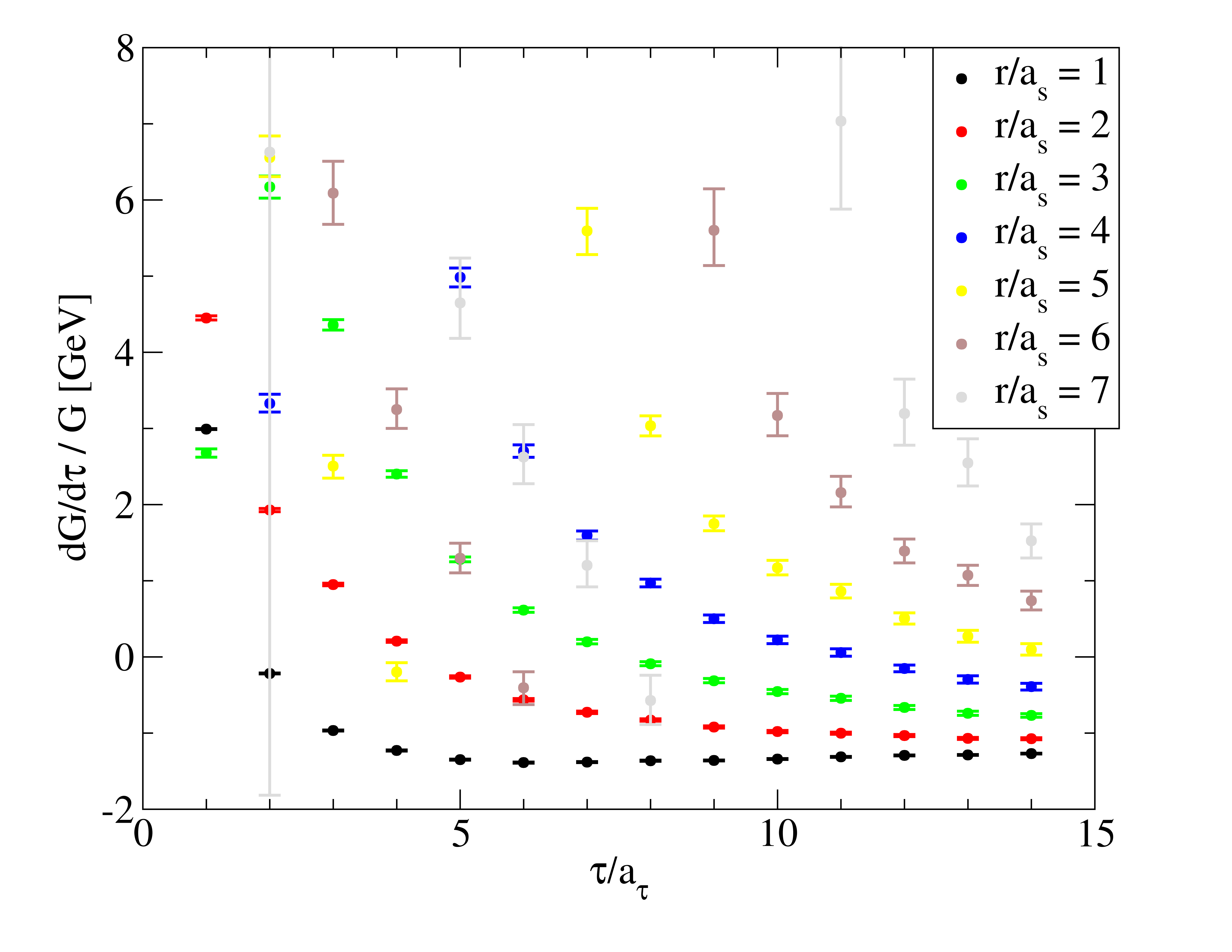}
\end{subfigure}
    \caption{The temporal derivative contribution to the $\Upsilon$ potential plotted against imaginary time for $T=141$MeV (left) and $T=352$MeV (right).
    }
    \label{fig:temporal_deriv_ntau40}
\end{figure}

In Fig. \ref{fig:temporal_deriv_ntau40} we show the temporal derivative term from
eq\eqref{eq:halqcd} for the same two temperatures.
This again shows the variation with $r$, although this is less than in the spatial
derivative case. We also note that there is a plateau at large $\tau$ visible for the $T=141$MeV case. This is to be expected, because at large $\tau$, the time derivative asymptotes to the ground state mass.

Comparing the two derivative from figs.\ref{fig:spatial_deriv_ntau40} \& \ref{fig:temporal_deriv_ntau40}, we see that the spatial derivative is numerically larger.

\subsection{Potentials for the $\eta_b$ and $\Upsilon$ Channels}

\begin{figure}[h]
\begin{subfigure}{.5\textwidth}
  \centering
    \includegraphics[width=\linewidth]{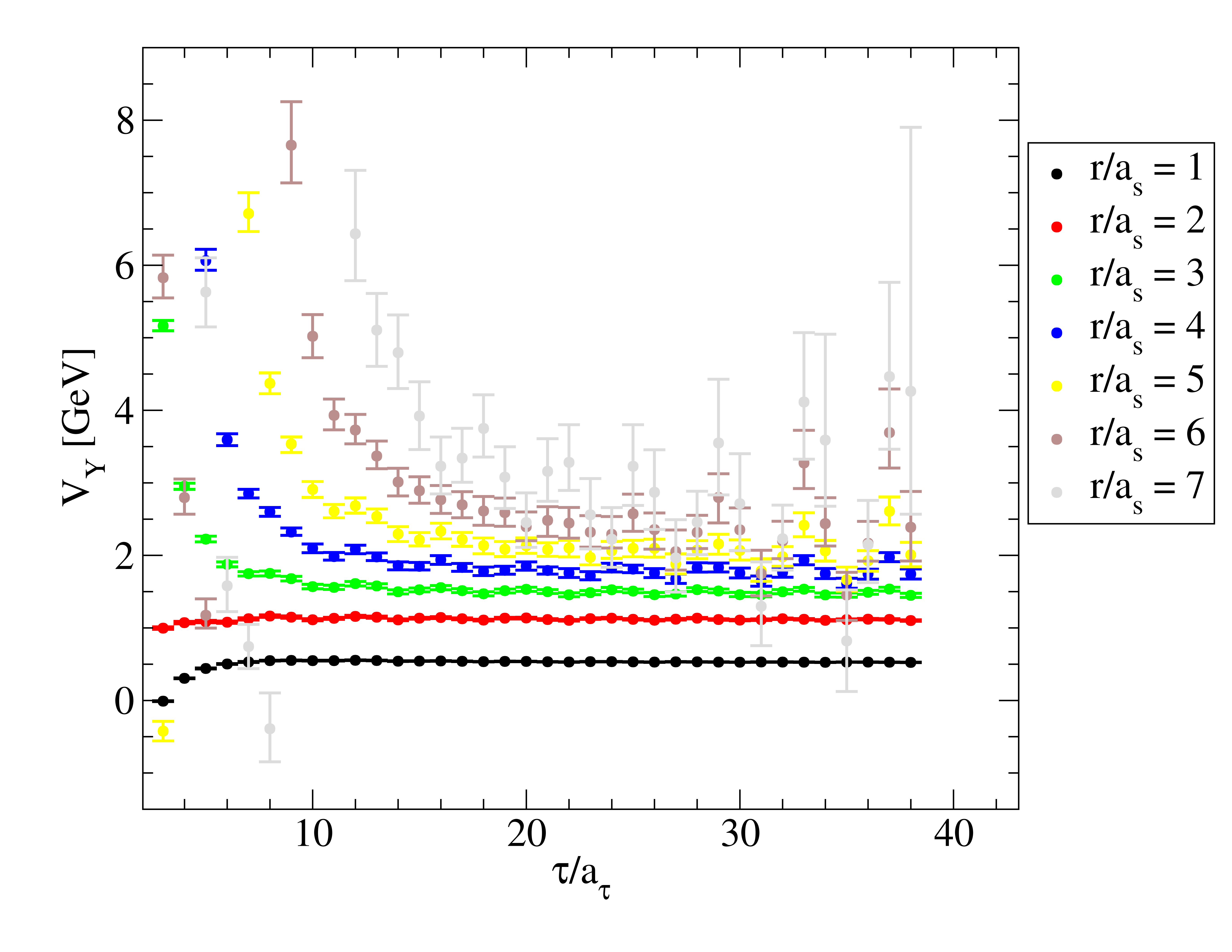}
\end{subfigure}
\begin{subfigure}{.5\textwidth}
  \centering
    \includegraphics[width=\linewidth]{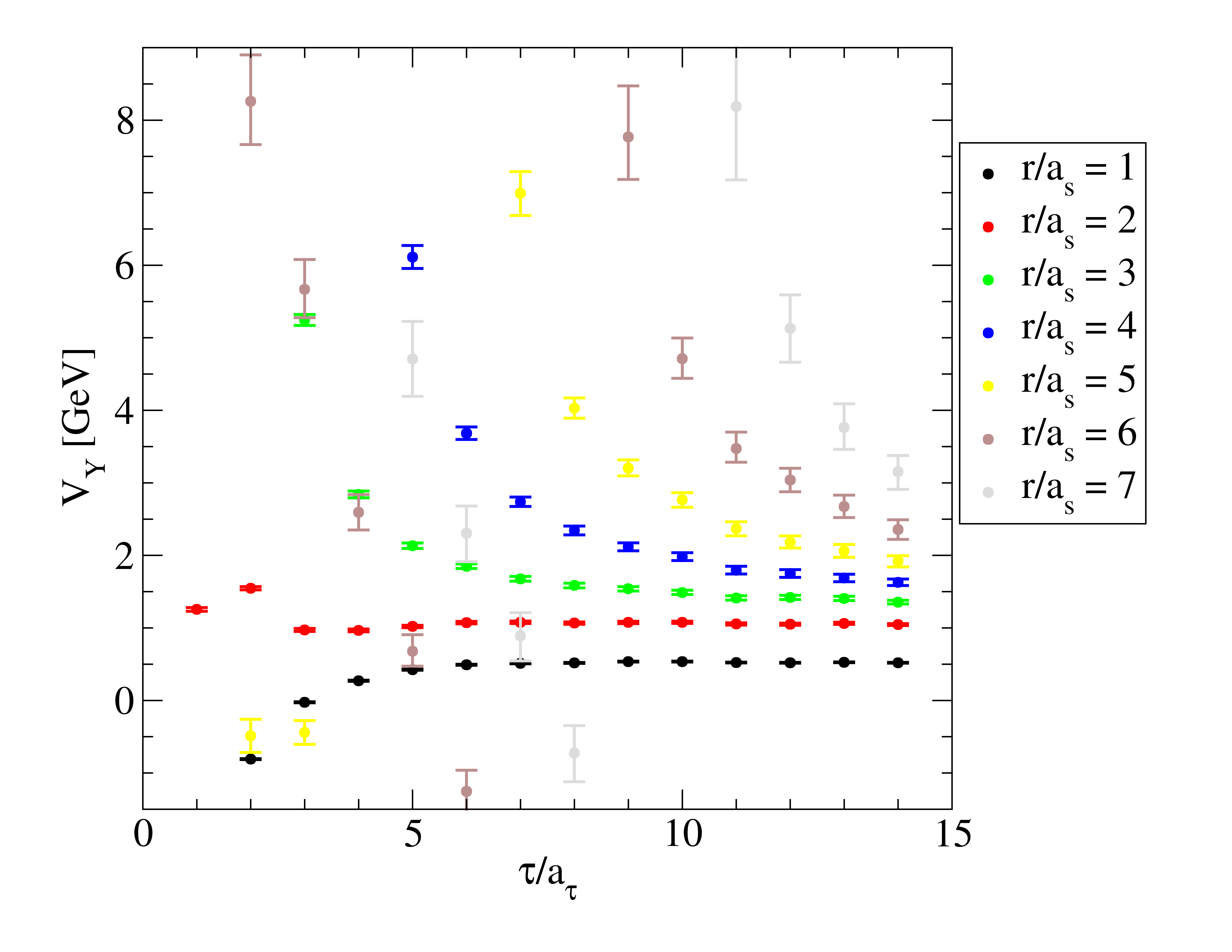}
\end{subfigure}
    \caption{The $\Upsilon$ potential plotted against imaginary time for $T=141$MeV (left) and $T=352$MeV (right).
    }
    \label{fig:V_tau}
\end{figure}

We combine the spatial and temporal derivatives for the $\eta_b$ and $\Upsilon$ channels
in eq\eqref{eq:halqcd} to obtain the the potentials for those channels, $V_{\eta_b}$ and $V_\Upsilon$.
The $\Upsilon$ case is plotted in fig.\ref{fig:V_tau}.
Note $V_{\eta_b}$ and $V_\Upsilon$ are explicit functions of $\tau$
in the time-dependent HAL QCD method,
due to the way these potentials are derived.
Ideally, they should be constant functions w.r.t. $\tau$, but as can be seen from
fig.\ref{fig:V_tau}, this is not the case, except at large $\tau$, or for small $r$.
We will investigate this $\tau$-dependency in future work by considering
lattice derivatives which are more sophisticated than those in
eqs\eqref{eq:s-der} \& \eqref{eq:t-der}.
We note that the results have the smallest systematics for small $r$ values.
We obtain our final estimate of the each channel's potential by averaging
over a time window as discussed in the next section.

\subsection{Central Potential}\label{sec:potential}

\begin{figure}[h]
    \centering
    \includegraphics[width=\textwidth]{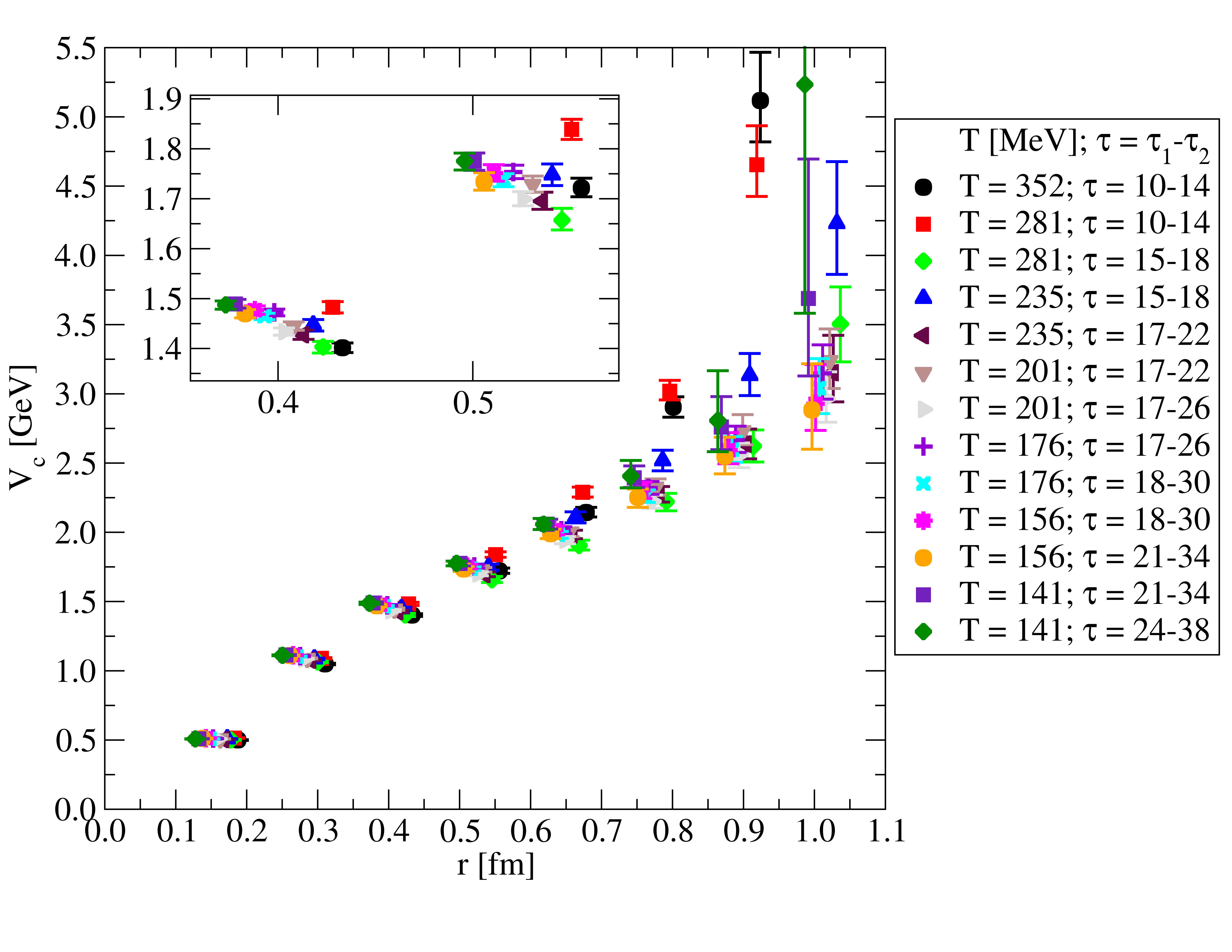}
    \caption{The central interquark potential in the bottomonium system plotted against quark separation $r$ for a range of temperatures. The points are offset horizontally for clarity. The potentials for each temperature are obtained by averaging two time ranges $[\tau_1,\tau_2]$ as indicated in the legend, chosen so that they are identical for neighbouring temperatures. This allows thermal effects to be disentangled from fitting systematics as discussed in the text.
    The insert shows a closeup of two $r$ values which indicates a thermal effect.
    }
    \label{fig:V_central}
\end{figure}

We average the potentials for the $\eta_b$ and $\Upsilon$ channels over the
time interval $\tau \in [\tau_1,\tau_2]$ (as discussed below) and then combine them
to obtain the central potential, $V_C(r)$ using eq\eqref{eq:Vcentral}.
Figure \ref{fig:V_central} plots $V_C(r)$ for all temperatures studied
(see Table \ref{tab:fastsum_details}).
Due to the periodic boundary conditions in the spatial direction,
there are only 13 distinct lattice points in the spatial direction.
However, the noise grows too quickly for points $r>7a_s$ to be considered.

Our aim is to determine the interquark potential as a function of temperature.
In order to disentangle possible systematic effects from thermal effects,
we use the {\em same} time window $[\tau_1,\tau_2]$ for neighbouring temperatures.
This ensures that the fitting procedure is identical for both of these temperatures,
and so any variation in the potential can be ascribed to a thermal, rather than systematic effect.
The data plotted in fig.\ref{fig:V_central} follows this procedure.
The points are off-set horizontally for clarity.
We see that the potential at large distances $r\gtrsim 0.8$fm have
large errors and become unstable for the hottest temperatures,
but those for distances $r \lesssim 0.8$fm have more modest errors
and are predictive.

From the plot, we see indications of thermal effects which are best seen in the insert
for the two distances $r\approx 0.4$ \& 0.5fm.
When considering neighbouring temperatures which share the same time window
$[\tau_1,\tau_2]$ we see that there is a clear trend towards a flattening of $V_C$
as $T$ increases.
This confirms our expectation that the interquark potential is temperature
dependent and becomes weaker with increasing temperature.

\section{Summary}

These proceedings present a calculation of the thermal interquark potential in the 
bottomonium system using the HAL QCD method with NRQCD quarks.
Our {\sc fastsum} Collaboration's anistropic "Generation 2" ensembles were used.
We find indications of thermal effects in the central potential, $V_C(r)$,
observing the expected flattening of the potential as the temperature increases.
Future work will use lattice derivatives which have smaller discretisation errors,
and a momentum space approach which will allow us to calculate the potential
at all spatial displacements.
We will also study higher order terms in the potential.

\section*{Acknowledgements}

This work is supported by STFC grant ST/T000813/1.
SK is supported by the National Research Foundation of Korea under grant NRF-2021R1A2C1092701 funded by the Korean government (MEST).
This work used the DiRAC Extreme Scaling service at the University of Edinburgh, operated by the Edinburgh Parallel Computing Centre on behalf of the STFC DiRAC HPC Facility (www.dirac.ac.uk). This equipment was funded by BEIS capital funding via STFC capital grant ST/K000411/1, STFC capital grant ST/H008845/1, and STFC DiRAC Operations grants ST/K005804/1 and ST/K005790/1. DiRAC is part of the National e-Infrastructure.
This work was performed using PRACE resources at Cineca via grants 2011040469 and 2012061129.
We acknowledge the support of the Supercomputing Wales project, which is part-funded by the European Regional Development Fund (ERDF) via Welsh Government.

\bibliographystyle{JHEP}
\bibliography{ref.bib}

\end{document}